\documentclass{elsart}
\usepackage{graphicx}
\usepackage{natbib}
\usepackage{amssymb}
\journal{Optics Communications}

\def\be{\begin{equation}}
\def\ee{\end{equation}}
\def\e#1{\label{#1}\end{equation}}
\def\bea{\begin{eqnarray}}
\def\eea{\end{eqnarray}}
\def\ea#1{\label{#1}\end{eqnarray}}

\def\bem#1{\begin{mathletters}\label{#1}}
\def\eml{\end{mathletters}}

\def\ket#1{{|#1\rangle}}
\def\bra#1{{\langle#1|}}
\def\braket#1#2{{\langle#1|#2\rangle}}
\def\4#1{{\bf{#1}}}
\def\8#1{{\widetilde{#1}}}
\def\eqref#1{(\ref{#1})}

\begin{document}
\begin{frontmatter}

\title{Universal Dynamical Control of Local Decoherence for Multipartite and Multilevel Systems}

\author{G. Gordon,}
\author{G. Kurizki\corauthref{cor}}
\corauth[cor]{Corresponding author.}
\ead{gershon.kurizki@weizmann.ac.il}
\and
\author{A. G. Kofman\thanksref{kof-address}}
\address{%
Department of Chemical Physics, Weizmann Institute of Science,
Rehovot 76100, Israel
}%

\thanks[kof-address]{%
On leave at the Dept. of Elec. Eng., UC, Riverside, CA 92521}

\begin{abstract}
A unified theory is given of dynamically modified decay and
decoherence of field-driven multilevel multipartite entangled
states that are weakly coupled to zero-temperature baths or
undergo random phase fluctuations. The theory allows for arbitrary
local differences in their coupling to the environment. Due to
such differences, the optimal driving-field modulation to ensure
maximal fidelity is found to substantially differ from
conventional ``Bang-Bang'' or $\pi$-phase flips of the
single-qubit evolution.
\end{abstract}

\end{frontmatter}

\section{Introduction}
\label{sec-intro}
A quantum system may decohere, under the influence of its
environment, in one (or both) of the following fashions: (a) Its
population may decay to a continuum or a thermal bath, a process
that characterizes spontaneous emission of photons by excited
atoms \cite{coh92}, vibrational and collisional relaxation of
trapped ions \cite{sac00} and cold atoms in optical lattices
\cite{gre02}, as well as the relaxation of current-biased
Josephson junctions \cite{cla88,leg87}. (b) It may undergo proper
dephasing, which randomizes the phases but does not affect the
population of quantum states, as in the case of phase interrupting
collisions \cite{scu97}.

Most theoretical and experimental methods aimed at assessing and
controlling (suppressing) the effects of decoherence of qubits
(any two-level system, that is the quantum equivalent of a
classical bit) have focussed on one of two particular situations:
(a) single qubits decohering {\em independently}; or (b) many
qubits {\em collectively} perturbed by the same environment. Thus,
quantum communication protocols based on entangled two-photon
states have been studied under collective depolarization
conditions, namely, {\em identical} random fluctuations of the
polarization for both photons \cite{ban04,bal04}. Entangled qubits
that reside at the same site or in equivalent sites of the system,
e.g. atoms in optical lattices, have likewise been assumed to
undergo identical decoherence.

For independently decohering qubits, the most powerful approach
suggested thus far for the suppression of decoherence appears to
be the ``dynamical decoupling'' (DD) of the system from the bath
\cite{aga00,aga01,aga01a,vio98,shi04,vit01,fac01,fac04,zan03,vio03,uch02,kho05,sto01,fao04}.
The standard ``bang-bang'' DD, i.e. $\pi$-phase flips of the
coupling via strong and sufficiently frequent resonant pulses
driving the qubit \cite{vio98,shi04,vit01}, has been proposed for
the suppression of proper dephasing \cite{sea00}. Several
extensions have been suggested to further optimize DD under proper
dephasing, such as multipulse control \cite{uch02}, continuous DD
\cite{vio03} and concatenated DD \cite{kho05}. DD has also been
adapted to suppress other types of decoherence couplings such as
internal state coupling \cite{sto01} and heating \cite{vit01}.

Our group has proposed a universal strategy of approximate DD
\cite{kof00,kof01b,kof01,kof04,kof04a,kof04,kof01a,kof96,pel04}
for both decay and proper dephasing, by either pulsed or
continuous wave (CW) modulation of the system-bath coupling. This
strategy allows us to tailor the strength and rate of the
modulating pulses to the spectrum of the bath (or continuum) by
means of a simple universal formula. In many cases, the standard
$\pi$-phases ``bang-bang'' is then found to be inadequate or
non-optimal.

In the collective decoherence situation, it is possible to single
out decoherence-free subspaces (DFS) \cite{vio00}, wherein
symmetrically degenerate many-qubit states, also known as ``dark''
or ``trapping'' states \cite{scu97}, are decoupled from the bath
\cite{zan03,zan97,lid98,wu02}.

Entangled states of two or more particles, wherein each particle
travels along a different channel or is stored at a different site
in the system, may present more challenging problems insofar as
combatting and controlling decoherence effects are concerned: if
their channels or sites are differently coupled to the
environment, is their entanglement more fragile? Is it harder to
protect? To answer these fundamental questions, we develop a very
general treatment. The present treatment extends our previously
published single-qubit universal strategy
\cite{kof00,kof01,kof04,bar04,gor05} to {\em multiple entangled
multilevel systems (particles)} which are either coupled to partly
correlated (or uncorrelated) zero temperature baths or undergo
locally-varying random dephasing. Furthermore, it applies {\em to
any difference} between the couplings of individual particles to
the environment. This difference may range from the
large-difference limit of completely independent couplings, which
can be treated by the single-particle dynamical control of
decoherence via modulation of the system-bath coupling, to the
opposite zero-difference limit of completely identical couplings,
allowing for multi-particle collective behavior and
decoherence-free variables
\cite{fac04,zan03,zan97,lid98,wu02,lid99,fac02,una03,bri05}. The
general treatment presented here is valid anywhere between these
two limits and allows us to pose and answer the key question:
under what conditions, if any, is {\em local control} by
modulation, addressing each particle individually, preferable to
{\em global control}, which does not discriminate between the
particles?

We show that in the realistic scenario, where the particles are
differently coupled to the bath, it is {\em advantageous to
locally control each particle by individual modulation, even if
such modulation is suboptimal} for suppressing the decoherence for
the single particle. This local modulation allows synchronizing
the phase-relation between the different modulations and
eliminates the cross-coupling between the different systems. As a
result, it allows us to preserve the multipartite entanglement and
reduces the multipartite decoherence problem to the single
particle decoherence problem. Throughout the paper we show the
advantages of local modulation, over global modulation (i.e.
identical modulation for all systems and levels), as regards the
preservation of arbitrary initial states, preservation of
entanglement and the intriguing possibility of entanglement
increase compared to its initial value.

In section~\ref{sec-gen-form} we present the general formalism, in
terms of the systems, their couplings to the baths and the
modulation used. In section~\ref{sec-zero} we investigate in
detail the coupling to zero-temperature baths for multiple
multilevel systems, and focus on two specific examples, namely a
single multilevel system and a singly-excited collective
entangled-state of many two-level-systems (TLS). This is followed
in section~\ref{sec-proper} by a description of multiple TLS
(qubits) undergoing proper dephasing, with a specific example of
Bell states. A discussion of the results is given in
section~\ref{sec-conc}.

\section{General Formalism}
\label{sec-gen-form}
Our total system is composed of $M$ systems, each having a ground
state and $N_j$ excited states, $\ket{g}_j$ and $\ket{n}_j$,
respectively, where $j=1,...,M$. Each of the excited states of
each system has a different energy, $\omega_{j,n}$. The $M$
systems are coupled to a bath and are subject to proper dephasing.
Since the coupling to the bath may differ from one system to
another and for every excited level, each is modulated by a
different Stark shift $\hbar\delta_{j,n}(t)$ and a driving field
$V_{j,n}(t)$. The total Hamiltonian is the sum of the system (S),
bath (B) and interaction (I) Hamiltonians:
\bea
\label{H-total}
&H(t)=&H^{(S)}(t)+H^{(B)}+H^{(I)}(t),\\
\label{H-S}
&H^S(t)=&\sum_{j=1}^M H^S_j(t)\otimes_{j'\neq j} I_{j'},\\
\label{H-S-i}
&H^S_j=&\hbar\sum_{n=1}^{N_j}
\left(\omega_{j,n}+\delta_{j,n}(t)\right) \ket{n}_j\,{}_j\bra{n}
\nonumber\\&&+\hbar
\sum_{n=1}^{N_j}V_{j,n}(t)\left(\ket{n}_j\,{}_j\bra{g} +
H.c.\right),\\
\label{H-B}
&H^B=&\hbar\sum_k \omega_k\ket{k}\bra{k},\\
\label{H-I}
&H^I(t)=&\hbar\sum_{j=1}^M\sum_k\sum_{n=1}^{N_j}
[\tilde\epsilon_{j,n}(t)\mu_{k,j,n}
\left(\ket{n}\bra{g}\right)_j\ket{vac}\bra{k}\otimes I_{j'\neq j}
\nonumber\\
&&+ H.c.]
+\hbar\sum_{j=1}^M\sum_{n=1}^{N_j}\delta^r_{j,n}(t)\ket{n}_j\,{}_j\bra{n}
\otimes_{j'\neq j} I_{j'}
\eea
Here $I$ is the identity operator, $\tilde\epsilon_{j,n}(t)$ is
the time dependent modulation field of the $n^{th}$ excited level
of system $j$, $\mu_{k,j,n}$ is the coupling coefficient of the
$n^{th}$ excited level of system $j$ to the first excited state
$\ket{k}$ of the single bosonic reservoir mode k, and its proper
dephasing $\delta^r_{j,n}(t)$ is treated semiclassically. $H.c.$
are Hermitian conjugates. The system Hamiltonian includes the
system terms, as well as the modulation (i.e. Stark-shift and
driving field) terms. The decoherence effects (both coupling to
the bath and the proper dephasing) compose the interaction
Hamiltonian.

\begin{figure}[ht]
\centering\includegraphics[width=8.5cm]{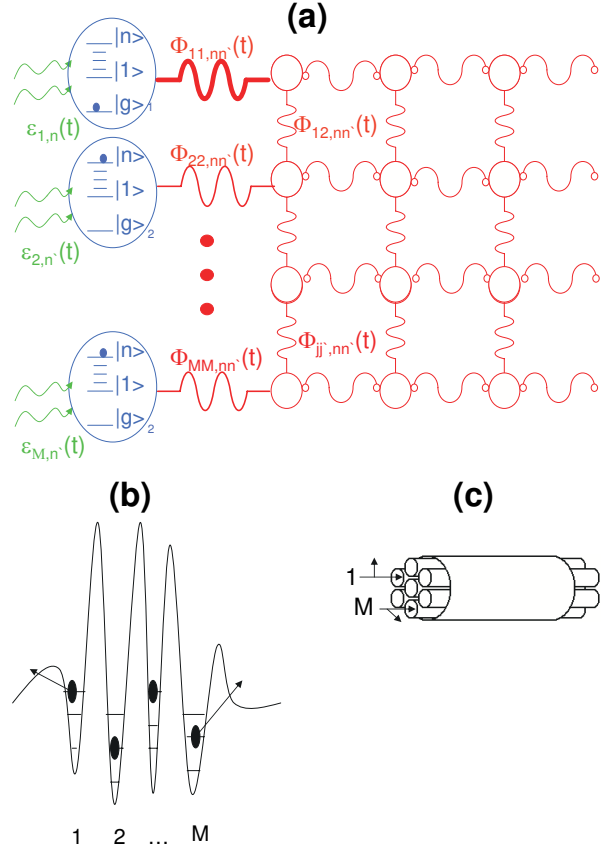}
 \caption{
 (a) Entangled multilevel systems with
 different couplings to a phonon bath or different proper dephasings, via
 $\Phi_{jj,nn'}(t)$. Their cross-coupling is through $\Phi_{jj',nn'}(t)$.
 The systems are modulated by $\epsilon_{j_n}(t)$.
 (b) Several polarization-entangled photons propagating through adjacent
 (coupled) fibers that exhibit fluctuating birefringence.
 (c) Entangled systems in tunnel-coupled multi-level wells of a washboard
 potential. There is no direct coupling between the wells at $t>0$,
 only different relaxation of each well to the continuum.}
 \protect\label{Fig-Schematic}
\end{figure}

Two decoherence scenarios will be discussed separately: (i) the
coupling to the zero-temperature bath is dominant and proper
dephasing is negligible, e.g. entangled atoms coupled to a cold
phonon bath; (ii) proper dephasing is dominant and one may neglect
the coupling to the bath, e.g. entangled photons in coupled
fluctuating birefringent fibers (Fig.~\ref{Fig-Schematic}). The
treatments of both scenarios are based on {\em analogous}
formalisms. The goal is to optimize the modulation or driving in
order to ensure maximal fidelity as time goes on.

The most general state in the system discussed here can be
represented in the basis of $N_T=\prod_{j=1}^MN_j$ states:
\be
\label{Psi-general}
\ket{\Psi}=\sum_{l=1}^{N_T} \lambda_l\ket{\Psi_l}
\ee
In quantum information (QI) implementations it is preferable to
use the interaction representation so that the fidelity of an
initial state, $\ket{\Psi_l}$, defined as
\be
\label{gen-fidelity}
F_l(t)=|\braket{\Psi_l}{\Psi(t)}|^2,
\ee
ensures that in the free (unperturbed) system it remains unity at
any time.

\section{Coupling to zero-temperature bath}
\label{sec-zero}
\subsection{General expressions for dynamical control of
zero-temperature decay}

First consider the scenario of different couplings of the systems
to a zero-temperature bath. The proper dephasing term is neglected
in eq.~\eqref{H-I}, i.e., $\delta^r_{j,n}(t)=0$, and we set the
driving fields to zero, i.e., $V_{j,n}(t)=0$.

The difference in the couplings to the k-th mode of the bath is
quantified by the cross product of their coupling coefficients,
(eq. \eqref{H-I}) $\mu_{k,j,n}\mu_{k,j',n'}$. Two extreme limits
can be discussed: (a) $\mu_{k,j,n}\mu_{k,j',n'}=0$ $\forall
k,j\neq j',n,n'$, in which the sets of $\{k_j\}$ modes are
separately coupled to each system, making the total hamiltonian
{\em separable} into contributions of the $M$ systems; (b)
$\mu_{k,j,n}=\mu_{k,j',n'}$ $\forall k,j,j',n,n'$, meaning that
the systems are identically coupled to the bath.

There is initially one excitation in our total system, thus the
full wave function in this scenario is:
\be
\label{zero-gen-full}
\ket{\Psi(t)}=\sum_k\alpha_0^k(t)\ket{k}\bigotimes_{j=1}^M\ket{g}_j
+\sum_{j=1}^M\sum_{n=1}^{N_j}\alpha_{j,n}(t)\ket{n}_j\ket{vac}
\bigotimes_{j'\neq j}\ket{g}_{j'}
\ee
where $\ket{vac}$ is the vacuum state of the bath. In order to
analyze the time-evolution of the wave function, written as a
column vector $\4\alpha(t)=\{\alpha_{j,n}(t)\}$, it is expedient
to express it in the interaction picture,
\be
\label{zero-gen-alpha}
\alpha_{j,n}(t)=e^{-i\omega_{j,n}t-i\int_0^td\tau\delta_{j,n}(\tau)}
\tilde{\alpha}_n(t).
\ee
The Schr\"{o}dinger equation for the coupled $\{\alpha_{j,n}(t)\}$
and $\{\alpha^k_0(t)\}$ amplitudes in \eqref{Psi-general} may be
reduced, upon eliminating the $\{\alpha^k_0(t)\}$ amplitudes and
transforming to the interaction picture, to the following exact
integro-differential equation:
\be
\label{zero-gen-integrodiff}
\dot{\tilde\alpha}_{j,n}(t)=\int_0^tdt'
\sum_{j',n'}\Phi^D_{jj',nn'}(t-t')K^D_{jj',nn'}(t,t')
e^{i\omega_{j,n}t-i\omega_{j',n'}t'}\tilde\alpha_{j',n'}(t')
\ee
Here $\4\Phi^D(t)$ is the reservoir-response matrix, given by:
\bea
\label{zero-gen-Phi}
&\Phi^D_{jj',nn'}(t)=&\int d\omega
G^D_{jj',nn'}(\omega)e^{-i\omega
t},\\
\label{zero-gen-G}
&G^D_{jj',nn'}(\omega)=&\hbar^{-2}\sum_k\mu_{k,j,n}\mu^*_{k,j',n'}
\delta(\omega-\omega_k).
\eea
and $\4K^D(t,t')$ is the modulation matrix, given by:
\be
\label{zero-gen-K}
K^D_{jj',nn'}(t,t')=\epsilon^{D*}_{j,n}(t)\epsilon^D_{j',n'}(t'),
\ee
where
\be
\label{zero-gen-epsilon}
\epsilon^D_{j,n}(t)=\tilde\epsilon_{j,n}(t)e^{-i\int_0^td\tau\delta_{j,n}(\tau)}
\ee
accounts for both the modulation and the Stark shift.

Since we are interested in modulations yielding slowly varying
solutions, we can pull
$\tilde\alpha_{j',n'}(t')\approx\tilde\alpha_{j',n'}(t)$ out of
the integrand in eq.~\eqref{zero-gen-integrodiff}, thereby
reducing it to the differential equation:
\be
\label{zero-gen-diff}
\dot{\tilde{\4\alpha}} = -\4W^D(t)\tilde{\4\alpha}.
\ee
Here the dynamically-controlled decoherence matrix,
$\4W^D(t)=\{W^D_{jj',nn'}(t)\}$, is a convolution of the
modulation and response matrices:
\be
\label{zero-gen-WD}
W^D_{jj',nn'}(t)=\int_0^tdt'
\Phi^D_{jj',nn'}(t-t')K^D_{jj',nn'}(t,t')
e^{i\omega_{j,n}t-i\omega_{j',n'}t'}
\ee
The solution to eq.~\eqref{zero-gen-diff} is given by:
\be
\label{zero-gen-solution}
\tilde{\4\alpha}(t)=\mathrm{T_+}e^{-\4J^D(t')}\tilde{\4\alpha}(0)
\ee
where $\4J^D=\{J^D_{jj',nn'}\}$ and
\be
\label{zero-gen-J-def}
J^D_{jj',nn'}(t) = \int_0^tdt'
\int_0^{t'}dt''\Phi^D_{jj',nn'}(t'-t'')K^D_{jj',nn'}(t',t'')
e^{i\omega_{j,n}t'-i\omega_{j',n'}t''}
\ee

Equations~\eqref{zero-gen-diff}-\eqref{zero-gen-J-def} are the
{\em most general expressions possible for the decoherence of
multilevel, multipartite entangled state under dynamical control}
by modulation. In what follows we explore the consequences of
these expressions.

Two alternative control strategies may be conceived of. The first
one is that of global modulation, meaning the modulation is
identical for all systems, $\epsilon^D_{j,n}=\epsilon^D_{j',n'}$
$\forall j,j',n,n'$. In this case, the decoherence matrix
\eqref{zero-gen-WD} retains its off-diagonal elements and the different states mix.
The alternative strategy is that of local modulations, i.e.
$\epsilon^D_{j,n}\neq\epsilon^D_{j',n'}$ $\forall j,j',n,n'$. It
will be shown advantageous to equalize the rates of decay of all
systems $\{j\}$ and all levels $\{n\}$, and to {\em avoid their
mixing} by the decoherence. These requirements amount to
fulfilling the following conditions:
\bea
\label{zero-gen-conditions-a}
&&J^D_{jj',nn'}=0\quad\forall j\neq j'\,{\rm or}\, n\neq n'\\
\label{zero-gen-conditions-b}
&&\frac{
\exp[-J^D_{j'j',n'n'}(t)-i\int_0^tdt'\delta_{j',n'}(t')]}{
\exp[-J^D_{jj,nn}(t)-i\int_0^tdt'\delta_{j,n}(t')]}=1
\quad\forall j,j',n,n'
\eea
which means that
\bea
\label{zero-gen-cond-real}
&&\Re J^D_{jj,nn}(t)=\Re J^D_{11,11} \quad \forall j,n\\
\label{zero-gen-cond-imag}
&&\Im J^D_{jj,nn}(t)+\int_0^tdt'\delta_{j,n}(t') = \Im
J^D_{11,11}(t)+\int_0^tdt'\delta_{1,1}(t') \quad [\mathrm{mod}
\,\,2\pi]
\eea

These conditions imply that {\em different modulations must be
applied to each system}, in {\em all cases}, whether the systems
are coupled to the same bath or to different baths. Our ability to
fulfil these conditions and, at the same time, minimize the
decay/decoherence of amplitudes $\alpha_{j,n}(t)$, can be
quantified in terms of the mixing $c_{j,n}$ and decay $A(t)$
parameters:
\bea
\label{zero-gen-mixing-parameter}
&&c_{j,n}(t)=\alpha_{j,n}(t)/\alpha_{1,1}(t) \\
\label{zero-gen-decay-parameter}
&&A(t)=\alpha_{1,1}(t)\sqrt{\sum_{j,n}|c_{j,n}(t)|^2}.
\eea

If only condition \eqref{zero-gen-conditions-a} is met, then
eq.~\eqref{zero-gen-mixing-parameter} yields
\be
\label{zero-gen-mixing}
c_{j,n}(t)=\frac{
\exp[-J^D_{jj,nn}(t)-i\int_0^tdt'\delta_{j,n}(t')]}{
\exp[-J^D_{11,11}(t)-i\int_0^tdt'\delta_{1,1}(t')]}
c_{j,n}(0).
\ee

In what follows we distinguish between two possible objectives:(i)
the preservation of the initial multipartite entangled state; (ii)
the steering of a partly-entangled (or unentangled) initial
multipartite state to a fully multipartite entangled state, both
in the presence of decoherence and modulation.

\subsection{Preservation of an initial entangled state}
If one wishes to {\em preserve an initial entangled state}, then
one should impose condition \eqref{zero-gen-conditions-b}, whereby
the different states {\em do not mix}, i.e.
$c_{j,n}(t)=c_{j,n}(0)$, but rather decay at a modified rate,
$J^D(t)=J^D_{11,11}(t)$, where
\be
\label{zero-gen-preservation-final}
\ket{\Psi(t)}=e^{-J^D(t)-i\int_0^tdt'\delta_{1,1}(t')}\ket{\Psi(0)}\\
\ee
The fidelity under these conditions is {\em identical} for all
initial states and is given by:
\be
\label{zero-gen-preservation-fidelity-local}
F_l(t)=e^{-2\Re J^D(t)}
\ee
Expression \eqref{zero-gen-preservation-final}, obtained under
conditions
\eqref{zero-gen-conditions-a}-\eqref{zero-gen-conditions-b}, is our
result for the optimal fidelity of preservation
\eqref{zero-gen-preservation-fidelity-local} under zero-temperature
decay: namely, optimal preservation requires the elimination of
state mixing and equal suppression of the decay for all systems.

\subsection{Steering}
If one wishes to {\em steer an initial state} $\ket{\Psi(0)}$, to
a desired state $\ket{\Psi^d}$, it is possible to exploit the
local modulation and the different decoherence rates in order to
acquire the desired state at a specific time $t$. The resultant
fidelity is then defined as
${F}^d(t)=|\braket{\Psi^d}{\Psi(t)}|^2$. In order to effectively
control the amplitude ratios of the states, while avoiding their
undesired mixing, it is expedient to define the mixing parameters
as
\be
c_{j,n}(t)=\alpha_{j,n}(t)/\alpha_{(j,n)_{\max}}(t)
\ee
where $(j,n)_{\max}$ is chosen such that $\alpha_{(j,n)_{\max}}$
is the largest amplitude of the desired state. This choice ensures
that each system $j$ and each level $n$ are controlled
independently (locally) without affecting the other systems. If
$(j,n)_{\max}=(1,1)$, then condition
\eqref{zero-gen-conditions-a} yields
\be
\label{zero-gen-dist-condition}
\frac{
\exp[-J^D_{11,11}(t)-i\int_0^tdt'\delta_{1,1}(t')]}{
\exp[-J^D_{jj,nn}(t)-i\int_0^tdt'\delta_{j,n}(t')]}
= \frac{c_{j,n}(0)}{{c^d}_{j,n}},
\ee
and at time $t$ the state has evolved to:
\be
\label{zero-gen-dist-psi}
\ket{\Psi(t)}=e^{-J^D(t)-i\int_0^tdt'\delta_{1,1}(t')}
\frac{\alpha_{1,1}(0)}{{\alpha^d}_{1,1}}\ket{\Psi^d}
\ee
where $J^D(t)=J^D_{11,11}(t)$ is given in
eq.~\eqref{zero-gen-J-def}. The resulting fidelity is
\be
{F^d}(t)=e^{-2\Re J^D(t)}|\alpha_{1,1}(0)/{\alpha^d}_{1,1}|^2.
\ee
Expression \eqref{zero-gen-dist-psi}, obtained under conditions
\eqref{zero-gen-conditions-a},\eqref{zero-gen-dist-condition}, yields
our result for the optimal steering towards entanglement fidelity
under zero-temperature decay: matching the decay rates and initial
mixing parameters to the desired mixing parameters.

Next, the implications of the foregoing general recipes for state
fidelity preservation and steering will be analyzed for the
following scenarios:

\subsection{Example: A single multilevel system}
In this case there is a single system with $N$ excited levels,
thus the $j$ subscript will be omitted. The bath response matrix
will be taken to be:
\be
\label{zero-Nlevel-Phi}
\Phi^D_{nn'}(t)=c_{nn'}d_n^*d_{n'}e^{-t^2/4t_c^2}
\ee
where $c_{nn'}$ is a constant coupling matrix $d_n=\cos\eta_n$,
with $\eta_n$ being the angle of transition dipole, and $t_c$ is
the correlation time of the bath. Here, the system eigenstates are
equidistant, $\omega_n=\omega_1+(n-1)\Delta$. We shall use
impulsive phase modulation,
$\epsilon^D_n(t)=e^{i[t/\tau]\theta_n}$. Here $[...]$ denote the
integer part, $\tau$ is the pulse duration and $\theta_n$ is the
phase change of level $n$.

\begin{figure}[htb]
\centering\includegraphics[width=8.5cm]{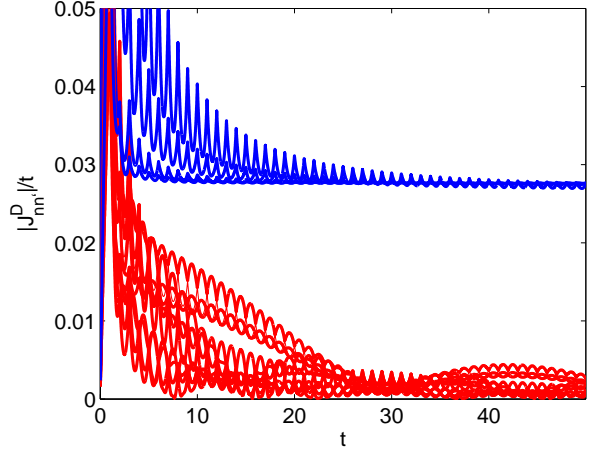}
 \caption{ $\4J^D$ as a function of time. Here $\omega_1=0.5$,
 $\Delta=0.1$, $t_c=1$,
 $\eta_n/\pi=\{0.246, 0.0, 0.326, 0.370\}$, and $c_{nn}=1.0, c_{nn'}=0.5$.
 The modulation interval time is $\tau=t_c$ and the phase changes are
 $\theta_n/\pi=\{1.0, 9.0, 8.0, 7.0\}$.
 The blue lines denote the 4 diagonal elements of $J^D$, while the red lines are the off-diagonal ones.}
\label{Figure-Multi-sym}
\end{figure}

In figure~\ref{Figure-Multi-sym} one can observe the
symmetrization of $J^D_{nn'}(t)$ as a function of time. The system
has $N=4$ levels. By choosing $\theta_n$ such that the long-time
limit of $J^D_{nn}$ is the same for all levels, one achieves the
elimination of the off-diagonal terms by different modulations,
and the symmetrization of the diagonal elements.

\begin{figure}[htb]
\centering\includegraphics[width=8.5cm]{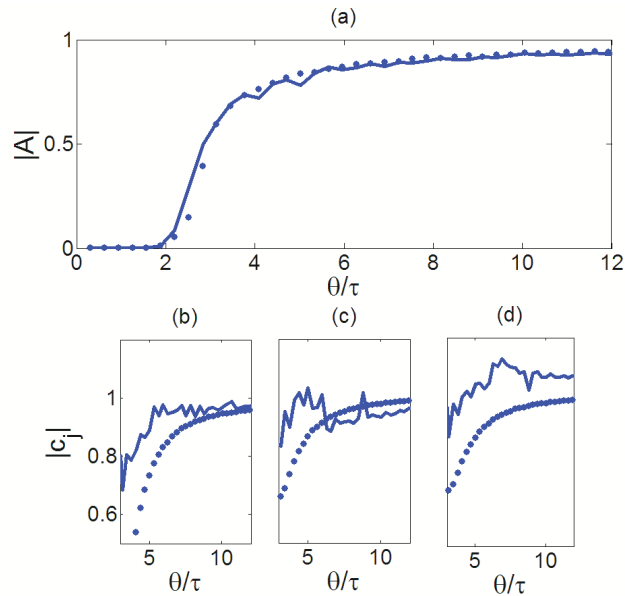}
\caption{Decay parameter $|A|$ (a) and mixing parameters
$|c_{2,3,4}|$ (b,c,d respectively) as a function of power invested
in impulse phase modulation. For global modulation (dotted)
$\theta=\pi$ whereas for local modulation (solid) for each $\tau$
the modulation phase $\theta_n$ is chosen such that symmetrization
is achieved (whenever possible). Here $t=50$ and other parameters
are as in figure \ref{Figure-Multi-sym}.}
\label{Figure-Multi-f}
\end{figure}

Figure~\ref{Figure-Multi-f} displays the decay and mixing
parameters as a function of the power $\hbar\theta_n/\tau$
invested in the impulse phase modulation. For global modulation,
$\theta_n=\pi$ is the optimized modulation phase for each level
coupled to a Gaussian bath, whereas for local modulation, the
phase modulation for each level $\theta_n$ at a given $\tau$ is
found such that symmetry is achieved, when possible. Due to the
simplicity of the modulation and the large differences in the
coupling to the bath, symmetrization was not always possible. The
$x$-axis units are those of the mean
$\theta/\tau=(1/N)\sum_n\theta_n/\tau$. As can be seen, one does
not increase the decay by using local modulation compared to
global modulation. However, whenever symmetrization is possible,
local modulation achieves greater preservation.

\subsection{Example: Entangled states of $M$ qubits}
Consider a system composed of $M$ two level systems (TLS) or
qubits, with ground and excited states, $\ket{g}$ and $\ket{e}$,
respectively. Only single-excitation states are considered here,
so the full wave function is given by :
\bea
\label{dec-M-TLS-Psi}
&\ket{\Psi}=\bigotimes_{j=1}^M\ket{g}_j + \sum_{l=1}^Md_l\ket{D^M_l}\\
\label{dec-M-TLS-Bell-basis}
&\ket{D^M_l}=\sum_{j=1}^Mq_j^{(l)} e^{i\omega_0
t}\ket{e}_j\bigotimes_{j'\neq j}\ket{g}_{j'}
\eea
where $\ket{D^M_l}$ provide a (completely entangled) basis for all
possible single-excitation states, $q_j^{(l)}=e^{2\pi
ij(l-1)/M}/\sqrt{M}$. $\ket{D^M_l}$ are zero sum amplitude states,
except for $\ket{D^M_1}$, which is the symmetric (Dicke) state. We
assume here that the TLS have the same excitation energies,
$\hbar\omega_0$.

Using the definitions in
eqs.~\eqref{zero-gen-mixing-parameter}-\eqref{zero-gen-decay-parameter}
of the decay and mixing parameters, the fidelity of an initial
entangled-state is a product of two terms, one due to decay and
the other due to mixing with other basis states:
\be
\label{dec-F}
F_l(t)=|A(t)|^2\frac{|\sum_{j=1}^Mq_j^{(l)}c_j(t)|^2}{\sum_{j=1}^M|c_j(t)|^2}.
\ee

One can maximize the fidelity by either reducing the decay, or
reducing (or even eliminating) the {\em mixing} of the
entangled-states. For the latter, one has to diagonalize and
equalize the diagonal elements of the decoherence matrix,
fulfilling the conditions in
eqs.~\eqref{zero-gen-conditions-a}-\eqref{zero-gen-conditions-b}.
If these conditions are met, the singly-excited entangled-states
{\em do not mix} with each other.

In the following numerical example, $M=3$ TLS are initially
prepared in a symmetric singly-excited Dicke state and are coupled
to a zero-temperature bath. The bath response matrix is taken to
be $\Phi^D_{jj'}(t)=\gamma
\frac{e^{-t^2/4t_j^2}e^{-t^2/4t_{j'}^2}} {r_0+r_{jj'}}$ where
$\gamma$ is a coupling constant, $t_j$ is the correlation time of
TLS $i$, $r_0$ is an arbitrary distance and
$r_{jj'}=|\4r_j-\4r_{j'}|$, where $\4r_j$ is the position of
system $j$. This model may describe residual absorption and
scattering (out of their initial modes) of three
polarization-entangled photons in adjacent nearly-overlapping
fibers \cite{gha98} or entangled, vibrationally relaxing atoms at
three inequivalent adjacent traps or lattice sites, all coupled to
the same continuum (fig.~\ref{Fig-Schematic}) \cite{fol05}.
Impulsive phase modulation is used as before,
$\epsilon^D_j(t)=e^{i[t/\tau]\theta_j}$.

\begin{figure}[htb]
\centering\includegraphics[width=8.5cm]{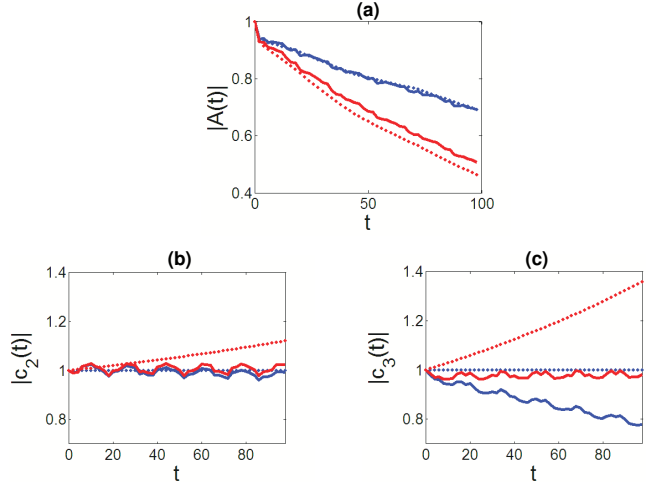}
 \caption{Entangled-state preservation. (a) Decay parameter $|A(t)|$ as a function of time. (b,c)
 Mixing parameters $|c_{2,3}(t)|$ as a function of time.
 The response matrix parameters are $\gamma=0.05$,
 $r_0=1.0$, the system parameter is $\omega=0.5$ and the modulation parameter
 is $\tau=1.0$.  The dotted (solid) lines indicate global (local) modulation,
 with $\theta_j/\pi=1.0$ ($\theta_j/\pi=\{1.0, 0.70, 0.58\}$) .  The blue (red) lines are for identical
 (different) coupling to the baths for the two TLS, with $t_j=1.0$
 ($t_j=\{0.75, 0.81, 1.0\}$).}
 \protect\label{Fig-dec}
\end{figure}

Figure~\ref{Fig-dec} shows the mixing and decay parameters
(eq.~\eqref{dec-F}) as a function of time, for three TLS with
cross-coupling between their relaxations, with $\4r_j =
\{r_0\cos(2\pi j/M),r_0\sin(2\pi j/M),0.0\}$. For an identical
coupling of all qubits (TLS) to the bath, i.e. equal correlation
times $t_j=t_{j'}$, one sees that a global modulation, meaning the
same modulation for the three TLS, results in zero mixing, whereas
local, or different, modulation results in increasing mixing with
time. However, for the case of different couplings to the bath,
i.e. $t_j\neq t_{j'}$, local modulation can eliminate the mixing,
if we choose the optimal modulation that equalizes the diagonal
decoherence matrix elements, whereas global modulation results in
an increased mixing with time.

For any difference in the coupling of the qubits, the results are
qualitatively similar for all initial singly-excited
entangled-states $\ket{D^M_l}$. Thus, for identical couplings,
local modulation achieves similar decay with increased mixing,
whereas for different couplings, {\em local modulation reduces
both decay and mixing} compared to the known global modulation
(the so-called ``parity-kicks'', i.e. $\pi$-phase flips for all
four qubits\cite{aga01,aga01a,vit01,vio99}). The optimal recipe
is, then, to apply $M$ {\em synchronous} pulse sequences to the
$M$ qubits, but with {\em locally-adapted pulse areas}:
$\theta_j/\pi=\{1.0, 0.70, 0.58\}$ in the example of
fig.~\ref{Fig-dec}.

\begin{figure}[htb]
\centering\includegraphics[width=8.5cm]{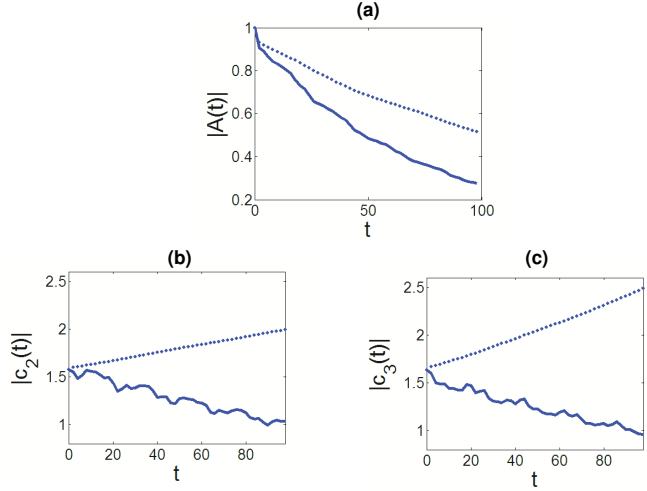}
 \caption{Steering towards a symmetric Dicke state. (a) Decay parameter $|A(t)|$ as a function of time. (b,c)
 Mixing parameters $|c_{2,3}(t)|$ as a function of time.
 The response matrix parameters are $\gamma=0.05$,
 $r_0=1.0$, $t_j=\{0.75, 0.81, 1.0\}$,
 the system parameter is $\omega=0.5$ and the modulation parameter is $\tau=1.0$.
 The dotted (solid) lines indicate global (local) modulation, with $\theta_j/\pi=1.0$
 ($\theta_j/\pi=\{0.80, 0.56, 0.47\}$).}
 \protect\label{Figure-dec-dist}
\end{figure}

Figure~\ref{Figure-dec-dist} shows the decay and mixing parameters
of steering an initial state which is a superposition of several
(completely entangled) basis states, with $c_j(0)=\{1.0, 1.57,
1.64\}$ and $A(0)=1.0$. The desired state is taken to be the
symmetric singly-excited Dicke state. One can see that using {\em
local modulation with different pulse rates} on the three qubits
causes the mixing parameters to approach their desired value (i.e.
${c^d}_{2,3}=1.0$). This result shows that one can {\em exploit
the different decoherence of the qubits in order to steer a
general initial state into any desired state}.

\section{Proper dephasing}
\label{sec-proper}
\subsection{Fidelity of $M$-qubit states}

Consider the scenario where the proper dephasing terms in
eqs.~\eqref{H-S-i},\eqref{H-I} are dominant, the coupling to the
bath and the Stark shifts being neglected. Here we assume from the
outset to have $M$ TLS (qubits) which undergo different proper
dephasings. The driving fields of each qubit, $V_j(t)$ are used to
dynamically reduce the proper dephasing.

In this case, the Hamiltonian is separable into parts pertaining
to the individual qubits. The wave function of any of them is
given by:
\be
\label{proper-gen-psi-i}
\ket\Psi_j=\beta_{jg}(t)\ket{g}_j+\beta_{je}(t)\ket{e}_j
\ee

Assuming, for simplicity, that the driving fields are resonant
with real envelope, i.e. $V_j(t)=V^{(0)}_j(t)e^{-i\omega_jt}
+c.c.$, one can change to the rotating frame by defining
$\tilde{\beta}_{je}(t)=e^{i\omega_jt}\beta_{je}(t)$ and
$\tilde{\beta}_{jg}(t)=\beta_{jg}(t)$ and use the rotating wave
approximation, neglecting terms oscillating at optical
frequencies. The diagonalizing basis of the system Hamiltonian of
each TLS is:
\be
\label{proper-gen-updown-def}
\ket{\uparrow}_j=\frac{1}{\sqrt{2}}
\left(e^{-i\omega_jt}\ket{e}_j+\ket{g}_j\right), \quad
\ket{\downarrow}_j=\frac{1}{\sqrt{2}}
\left(e^{-i\omega_jt}\ket{e}_j-\ket{g}_j\right)
\ee
Each TLS wave function is now described by
\be
\label{psi-pm-def}
\ket\Psi_j=\beta_{j+}(t)\ket{\uparrow}_j+\beta_{j-}(t)\ket{\downarrow}_j,
\ee
where
\be
\label{alpha-pm-def}
\beta_{j\pm}(t)=\frac{1}{\sqrt{2}}
\left(\tilde\beta_{je}(t)\pm\tilde\beta_{jg}(t)\right),
\ee
which results in the single-TLS dynamical equation:
\be
\label{proper-gen-beta-dot}
\dot{\beta}_{j\pm}(t)=\mp
iV^{(0)}_j(t)\beta_{j\pm}(t)-i\frac{\delta^r_j(t)}{2}
\left(\beta_{j+}(t)+\beta_{j-}(t)\right)
\ee

The full wave function is composed of $N_T=2^M$ basis states,
$\ket{\Psi_l}$, $l=1...N_T$, which can be presented in a binary
representation, meaning $l=b_1^lb_2^l...b_N^l$, with $b_j^l=0,1$.
Here zero denotes $\uparrow$ or plus sign and one denotes
$\downarrow$ or minus sign. Thus each basis state $\ket{\Psi_l}$
is a product of $M$ TLS states:
\be
\label{proper-gen-basis}
\ket{\Psi_l}=\bigotimes_{j=1}^M\ket{b_j^l}_j
\ee
and the full wave function is given by:
\be
\label{proper-gen-psi}
\ket{\Psi}=\sum_{l=1}^{N_T}\beta_l\ket{\Psi_l}
\ee
where
\be
\label{proper-gen-beta-k}
\beta_l=\prod_{j=1}^M\beta_{jb_j^l}
\ee

In order to solve for the wave function \eqref{Psi-general}, it is
useful to define the column vector $\4\beta=\{\beta_l\}$ and adopt
the matrix formulation. We next transform the column vector to
account for the driving fields:
\bea
\label{proper-gen-g-tilde}
&&\tilde{\4\beta}=e^{i\int_0^t dt' \4P(t')}\4\beta \\
\label{proper-gen-F}
&&P_{ll'}(t)=\delta_{ll'}\sum_{j=1}^M(2b_j^l-1)V_j(t)
\eea
where $\delta_{ll'}$ is Kronecker's delta. This vector fulfills
the following dynamical equation:
\be
\label{proper-gen-dyn}
\dot{\tilde{\4\beta}}=-(i/2)\4W^P(t)\tilde{\4\beta}
\ee
The transformed proper-dephasing matrix can be split into a part
that is proportional to the identity $\4I$ and an off diagonal
part:
\bea
\label{proper-gen-W-P}
&\4W^P(t)=&\4I\sum_{j=1}^M\delta^r_j(t)+\4W^{P,{\rm off}}(t),\\
\label{proper-gen-W-P-off}
&W^{P,{\rm off}}_{l,l'}(t)=&\delta_{|l-l'|_b,1}
\delta^r_{j_{l'\rightarrow l}}e^{is^{l'\rightarrow l}_j
\phi_{j_{l'\rightarrow l}}(t)}\\
\label{proper-gen-phi-i}
&\phi_j(t)=&2\int_0^t dt' V^{(0)}_j(t')
\eea
Here $|l-l'|_b$ is the binary distance between $l$ and $l'$,
measuring how many qubit-flips are required to get from $l'$ to
$l$, $j_{l'\rightarrow l}$ is the qubit required to flip in order
to get from $l'$ to $l$ and the sign function $s^{l'\rightarrow
l}_i=b^l_{j_{l'\rightarrow l}}-b^{l'}_{j_{l'\rightarrow l}}$ is
$+1$ for a qubit flip $1\rightarrow0$ and $-1$ for a qubit flip
$0\rightarrow1$. A qubit-flip of qubit $j$ means a change of
$\ket{\uparrow}_j\leftrightarrow\ket{\downarrow}_j$ (eq.
\eqref{proper-gen-updown-def}) and not the information qubit flip
$\ket{e}_j\leftrightarrow\ket{g}_j$.

The off-diagonal terms of the transformed proper-dephasing matrix
are non-zero only for elements which require a single qubit flip,
and are equal to the product of the proper-dephasing rate of the
qubit flipped and the modulation phase of that qubit, with the
appropriate sign $e^{\pm i\phi_{l\rightarrow
l'}(t)}\delta_{j_{l\rightarrow l'}}(t)$.

Since the proper dephasing term is stochastic, one must define the
first and second ensemble-averaged-moments, as
$\overline{\delta^r_j}(t)=0$ and
$\Phi_{jj'}^{P}(t)=\overline{\delta^r_j(t)\delta^r_{j'}(0)}$,
respectively, and adapt the solution to the density matrix
$\tilde{\4\rho}(t)=\overline{\tilde{\4\beta}(t)\tilde{\4\beta}^\dagger(t)}$.
Solution of eq.~\eqref{proper-gen-dyn} to second order in
$\delta^r_j$ then corresponds to:
\be
\label{proper-gen-solution}
\overline{\4\rho}(t)=\tilde{\4\rho}(0)-
\frac{1}{4}\int_0^tdt'\int_0^{t'}dt''
\overline{[\4W^P_{off}(t'),[\4W^P_{off}(t''),\tilde{\4\rho}(0)]]}
\ee
It describes the evolution of the density matrix under two
consecutive (virtual) qubit flips, i.e. excitation and
deexcitation, ending up with the same number of excitations, but
with a random (stochastic) phase.

The fidelity of an initial basis state,
$\4\rho_k(0)=\ket{\Psi_l}\bra{\Psi_l}$ is now defined as:
\be
\label{proper-gen-fidelity}
F_l(t)=\bra{\Psi_l}\overline{\4\rho}(t)\ket{\Psi_l}
\ee
It is identical for all initial basis states and is found to be:
\bea
\label{proper-gen-F-k}
&&F(t)=1-\frac{1}{2}\sum_{j=1}^M \Re J^P_{jj}(t)\\
\label{proper-gen-J-def}
&&J^P_{jj'}(t)=\int_0^tdt'\int_0^{t'}dt''
\Phi^P_{jj'}(t'-t'')K^P_{jj'}(t',t'')\\
\label{proper-gen-K}
&&K^P_{jj'}(t',t'')=\epsilon^{P*}_j(t')\epsilon^P_{j'}(t'')
\eea
where $\epsilon^P_j(t)=e^{i\phi_j(t)}$. Since the basis states are
product states, this fidelity does not pertain to entanglement.
Different modulations of individual qubits do not affect it. In
order to explore the implications of local and global modulations,
we shall revert to the basis of entangled Bell states.

\subsection{Fidelity control of Bell states under proper dephasing}

\subsubsection{General recipe}
Let us take two TLS, or qubits, which are initially prepared in a
Bell state. We wish to obtain the conditions that will preserve
it. In order to do that, we revert to the Bell basis, which is
given by
\bea
\label{dec-two-TLS-Bell-basis}
&&\ket{B_{1,2}}=1/\sqrt{2}e^{i\omega_0t}\left(\ket{e}_1\ket{g}_2
\pm
\ket{g}_1\ket{e}_2\right)\\
\label{proper-bell-def}
&&\ket{B_{3,4}}=1/\sqrt{2}\left(e^{i2\omega_0t}\ket{e}_1\ket{e}_2
\pm \ket{g}_1\ket{g}_2\right).
\eea
This is done by applying the proper rotation matrix to
eq.~\eqref{proper-gen-solution}. For an initial Bell-state
$\overline{\4\rho}_l(0)=\ket{B_l}\bra{B_l}$, where $l=1...4$, one
can then obtain the fidelity,
$F_l(t)=\bra{B_l}\overline{\4\rho}_l(t)\ket{B_l}$, as:
\bea
\label{proper-two-TLS-F}
&F_{l}(t)=\cos(\phi_\pm(t))\Re\left[e^{i\phi_\pm(t)}
\left(1-\frac{1}{2}\sum_{jj'}J^P_{jj',l}(t)\right)\right]
\\
\label{proper-two-TLS-J-def}
&J^P_{jj',l}(t)=\int_0^tdt'\int_0^{t'}dt''
\Phi^P_{jj'}(t'-t'')K^P_{jj',l}(t',t'') \\
\label{proper-two-TLS-Lambda-k-2}
&K^P_{jj,l}(t,t')=\epsilon^{P*}_j(t)\epsilon^P_j(t')\\
\label{proper-two-TLS-Lambda-k-3}
&K^P_{jj',3}(t,t')=-K^P_{jj',1}(t,t')=\epsilon^{P*}_j(t)\epsilon^{P*}_{j'}(t')\\
\label{proper-two-TLS-Lambda-k-4}
&K^P_{jj',4}(t,t')=-K^P_{jj',2}(t,t')=\epsilon^P_j(t)\epsilon^{P*}_{j'}(t')
\eea
where $\phi_\pm(t)=(\phi_1(t)\pm\phi_2(t))/2$ and the $\phi_+$
corresponds to $k=1,3$ and $\phi_-$ to $k=2,4$.

Expressions
\eqref{proper-two-TLS-F}-\eqref{proper-two-TLS-Lambda-k-4} provide
our recipe for minimizing the Bell-state fidelity losses. They
hold for {\em any} dephasing time-correlations and {\em arbitrary}
modulation.

\subsubsection{Numerical example} In the next numerical example,
the response matrix is taken to be
\be
\label{proper-dicke-numerical-phi}
\Phi^P_{jj'}(t)=\gamma e^{-\frac{t}{2t_j}-\frac{t}{2t_{j'}} -
r^2_{jj'}}
\ee
where $\gamma$ is a coupling constant, $t_j$ is the correlation
time of TLS $i$, and $r_{jj'}=|\4r_j-\4r_{j'}|$, where $\4r_j$ is
the position of particle $j$. This model may again describe
multi-photon \cite{scu97} or multi-atom decoherence, as above
(fig. \ref{Fig-Schematic}). Impulsive phase modulation is used as
before $\epsilon^P_j(t)=e^{i[t/\tau_j]\theta_j}$.

\begin{figure}[hb]
\centering\includegraphics[width=8.5cm]{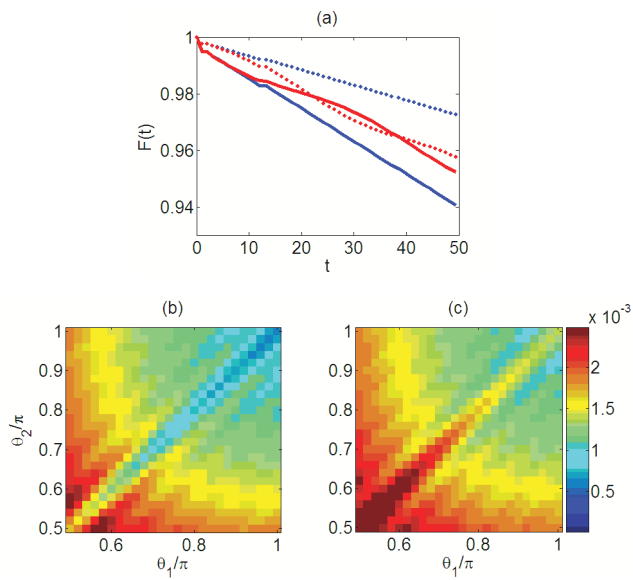}
 \caption{(a) Fidelity as a function of time. The blue (red) lines
 indicate triplet (singlet) initial states, whereas the solid
 (dashed) lines show the effects of global (local) modulation,
 with $\theta_2=0.9\pi$ ($\theta_2=0.8\pi$).
 (b,c) Dephasing rates of singlet (b) and triplet (c) initial Bell-states,
 as a function of modulation phases, $\theta_{1,2}$.
 The response matrix parameters are $\gamma=0.01$, $t_1=t_2=1.0$ and
 $r_0=1.0$ and the modulation parameters are
 $\tau_1=\tau_2=1.0$, $\theta_1=0.9\pi$.}
 \protect\label{Fig-pro-1}
\end{figure}

Figure~\ref{Fig-pro-1} shows the fidelity, $F_l(t)$ as a function
of time for the singlet, $\ket{B_2}$ and triplet, $\ket{B_4}$
states under global and local modulation. The global modulation is
seen to affect the different states in a different manner: the
singlet state decoheres more slowly than the triplet state.
However, the local (different) modulation for the different TLS,
eliminates the cross-coupling terms and equalizes the decoherence
rates of the two states. One can further see, by inserting the
aforementioned response matrix into
\eqref{proper-two-TLS-J-def}, that local modulation has the
same effect as the decorrelation of the two TLS-bath couplings,
i.e. each of the entangled qubits now decoheres independently, as
for $r_{12}\rightarrow\infty$.

Thus, locally induced decorrelation of the dephasings can either
reduce or enhance the Bell-state fidelity compared to standard
global (``Bang-Bang'') $\pi$-phase flips, depending whether the
correlated dephasings interfere constructively (for triplets) or
destructively (for singlets).

If we use singlet and triplet intermittently to encode
information, it is advantageous to use local modulation (in
fig.~\ref{Fig-pro-1}: $\theta_1=0.9\pi, \theta_2=0.8\pi$) to
equate their fidelities, rather than the standard ``Bang-Bang''.

\section{Conclusions}
\label{sec-conc}

In this paper we have expounded our comprehensive approach to the
dynamical control of decay and decoherence. Our analysis of
multiple field-driven multilevel systems which are coupled to
partly-correlated or independent baths or undergo locally-varying
random dephasing has resulted in the universal formulae
\eqref{zero-gen-diff}-\eqref{zero-gen-WD} for coupling to
zero-temperature bath and
\eqref{proper-gen-W-P-off}-\eqref{proper-gen-solution} for proper
dephasing. The merits of local vs. global modulations were
presented and are summarized below:
\begin{itemize}

\item For different couplings to a zero-temperature bath, one can
better preserve any initial state by using local modulation which
can reduce the decay as well as the mixing with other states, more
than global modulation. For a single multilevel system, it was
shown that local modulation which eliminates the cross-decoherence
terms, increases the fidelity more than the global modulation
alternative. For two TLS, it was shown that local modulation
better preserves an initial Bell-state, whether a singlet or a
triplet, compared to global $\pi$-phase ``parity kicks''.

\item One can exploit the different couplings to a
zero-temperature bath and local modulation in order to steer an
initial partly-entangled or unentangled state, to a desired
entangled multipartite state. The ability to match the decoherence
rates to the desired mixing parameters is made possible by using
local modulation, which results in lower fidelity losses compared
to global modulation.

\item Local modulation can effectively {\em decorrelate} the
different proper dephasings of the multiple TLS, resulting in
equal dephasing rates for all states. For two TLS, we have shown
that the singlet and triplet Bell-states acquire the same
dynamically-modified dephasing rate. This should be beneficial
compared to the standard global ``Bang-Bang'' ($\pi$-phase flips)
if both states are used (intermittently) for information
transmission or storage.
\end{itemize}

Our general analysis allows one to come up with an optimal choice
between global and local control, based on the observation that
the maximal suppression of decoherence is not necessarily the best
one. Instead, we demand an optimal {\em phase-relation} between
different, but {\em synchronous} local modulations of each
particle.

\ack We acknowledge the support of ISF and EC (ATESIT,QUACS and
SCALA Networks).

\newpage


\end{document}